\newcommand{\dunlap}{\supit{a}}
\newcommand{\ubc}{\supit{b}}
\newcommand{\mcg}{\supit{c}}
\newcommand{\cita}{\supit{d}}
\newcommand{\cifar}{\supit{e}}
\newcommand{\ut}{\supit{f}}
\newcommand{\ibm}{\supit{g}}
\newcommand{\drao}{\supit{h}}
\newcommand{\cmu}{\supit{i}}

\documentclass[12pt]{spie}
\usepackage[]{graphicx}
\usepackage{setspace}
\usepackage{tocloft}
\usepackage{color}
\usepackage[]{bm}
\usepackage[]{amsmath}
\usepackage{verbatim}
\usepackage{hyperref}
\usepackage[lofdepth,lotdepth]{subfig}

\title{Calibrating CHIME, A New Radio Interferometer to Probe Dark Energy} 
\author{ Laura B. Newburgh\dunlap,
Graeme E. Addison\ubc, 
Mandana Amiri\ubc, 
Kevin Bandura\mcg, 
J. Richard Bond\cita\cifar, 
Liam Connor\cita\ut, 
Jean-Fran\c{c}ois Cliche\mcg, 
Greg Davis\ubc, 
Meiling Deng\ubc, 
Nolan Denman\ut, 
Matt Dobbs\mcg, 
Mateus Fandino\ubc,
Heather Fong\ubc,
Kenneth Gibbs\ubc, 
Adam Gilbert\mcg, 
Elizabeth Griffin\ubc,
Mark Halpern\ubc, 
David Hanna\mcg, 
Adam D. Hincks\ubc, 
Gary Hinshaw\ubc, 
Carolin H\"ofer\ubc, 
Peter Klages\ut\ibm, 
Tom Landecker\drao, 
Kiyoshi Masui\ubc\cifar, 
Juan Mena Parra\mcg,  
Ue-Li Pen\cita, 
Jeff Peterson\cmu, 
Andre Recnik\ut, 
J. Richard Shaw\cita, 
Kris Sigurdson\ubc, 
Michael Sitwell\ubc, 
Graeme Smecher\mcg, 
Rick Smegal\ubc, 
Keith Vanderlinde\ut\dunlap, and
Don Wiebe\ubc
\skiplinehalf
\begin{small}
\supit{a}Dunlap Institute for Astronomy \& Astrophysics, University
of Toronto, 50 St George St, Toronto, ON, M5S 3H4, Canada
\skiplinehalf
\supit{b}Department of Physics and Astronomy, University of British Columbia, 
6224 Agricultural Rd. Vancouver, V6T 1Z1, Canada
\skiplinehalf
\supit{c}McGill University, 3600 University St, Montreal, Canada
\skiplinehalf
\supit{d}CITA, 60 St George St, Toronto, ON, M5S 3H8, Canada
\skiplinehalf
\supit{e}Canadian Institute for Advanced Research,
CIFAR Program in Cosmology and Gravity,
Toronto, ON M5G 1Z8
\skiplinehalf 
\supit{f}Department of Astronomy \& Astrophysics, University of
Toronto, 50 St George St, Toronto, ON, M5S 3H4, Canada
\skiplinehalf
\supit{g}IBM Canada
\skiplinehalf
\supit{h}National Research Council Canada, Dominion Radio Astrophysical
Observatory, Box 248, Penticton BC V2A 6J9 Canada
\skiplinehalf
\supit{i}McWilliams Center for Cosmology, Carnegie Mellon University,
Department of Physics, 5000 Forbes Ave, Pittsburgh PA 15213, USA
\skiplinehalf
\end{small}
}

\authorinfo{Send correspondence to newburgh@di.utoronto.ca}

\begin{document} 
\maketitle 

\begin{abstract}
The Canadian Hydrogen Intensity Mapping Experiment (CHIME) is a transit interferometer currently being built at the Dominion Radio Astrophysical Observatory (DRAO) in Penticton, BC, Canada. We will use CHIME to map neutral hydrogen in the frequency range 400 -- 800\,MHz over half of the sky, producing a measurement of baryon acoustic oscillations (BAO) at redshifts between 0.8 -- 2.5 to probe dark energy. We have deployed a pathfinder version of CHIME that will yield constraints on the BAO power spectrum and provide a test-bed for our calibration scheme. I will discuss the CHIME calibration requirements and describe instrumentation we are developing to meet these requirements.
\end{abstract}

\keywords{21cm, cosmology, calibration, dark energy}


\section{Introduction}
\label{sect:intro}
CHIME is a new radio interferometer currently being built at the Dominion Radio Astrophysical Observatory (DRAO) that will operate between 400 -- 800\,MHz. The primary scientific goal of CHIME is to constrain the dark energy equation of state by measuring the expansion history of the Universe through its impact on large-scale structure. We will measure the BAO comoving $\sim$150\,Mpc scale\cite{SDSS3:2014,SDSSLRG:2005,2dF:2005,WiggleZ:2011,SDSSlya:2013,BOSSlya:2013} in a redshift range $0.8 < z < 2.5$, which encompasses the epoch when the $\Lambda$CDM\cite{WMAP9:2013} model predicts dark energy began to dominate the energy density of the Universe. To obtain sufficient survey volume for this measurement, we must map large-scale structure over half of the sky. To achieve this goal, we will use the neutral hydrogen 21\,cm emission to trace the large-scale matter distribution over our required redshift range\cite{Chang:2010,Switzer:2013,Masui:2013}  across the entire sky accessible from the DRAO. 

The CHIME instrument consists of five adjacent $f/0.25$ cylindrical dishes, each 20\,m wide by 100\,m long and featuring an array of close-packed feeds along the focal lines. The stationary cylinders are oriented north-south to form a transit telescope as the sky rotates from east to west. The frequency-dependent resolution of the instrument is fixed by its longest baseline to be 0.26$^{\circ}$ -- 0.52$^{\circ}$. This resolution is optimized for measurements of BAO features in the hydrogen maps within our redshift range. 

\begin{figure}[h]
   \begin{center}
   \begin{tabular}{c}
  \includegraphics[width=0.7\textwidth]{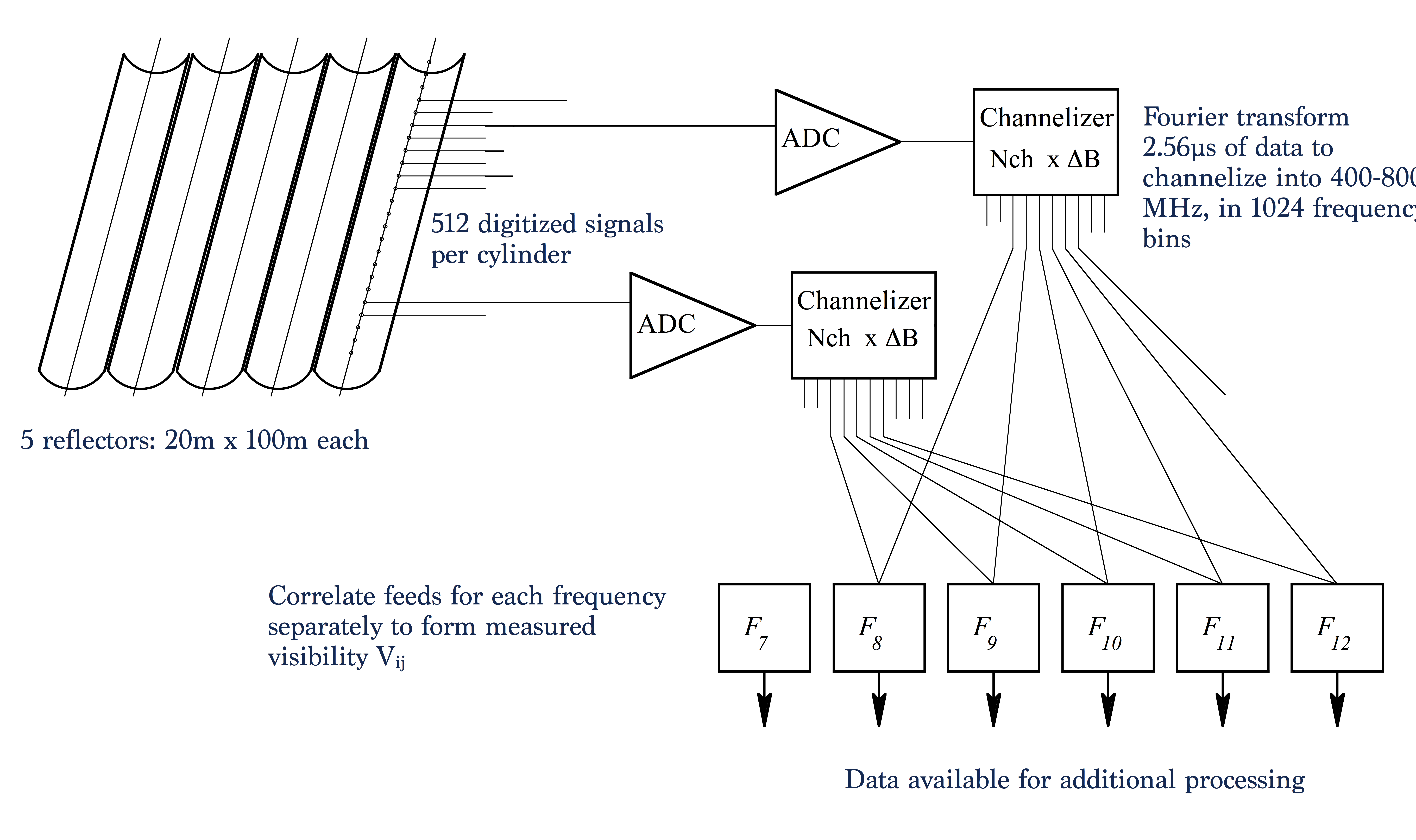}
     \end{tabular}
   \end{center}
   \caption{Schematic of the five cylindrical dishes and data processing which comprise the CHIME interferometer. There are 256 dual-polarization feeds on each cylinder whose signals are channelized into $\sim$0.3\,MHz bins, and the signals from pairs of feeds are correlated. }
   \label{fig:CHIMElayout} 
 \end{figure}

The sky signal is focused by the cylindrical dishes onto dual-polarization feeds fixed to the focal line of each cylinder. The resulting signals are amplified by ambient-temperature first-stage low-noise amplifiers (LNAs) attached directly to the output of the feeds. The signal is then carried down either coaxial cable or a radio frequency-over-fiber system to a second set of amplifiers and a 400 -- 800\,MHz bandpass filter. As shown in Figure~\ref{fig:CHIMElayout}, the signals are then digitized, channelized into 1024 frequency bands, and correlated. The feeds are spaced $\sim$30\,cm along the focal line, in a highly redundant pattern. The salient features of CHIME are summarized in Table~\ref{tab:salient}. 

\begin{table}[h]
\centering
\begin{tabular}[b]{| ll |} \hline\hline
\rule{0pt}{4ex}Location & DRAO ($49^{\circ}19^{\prime}$N, $119^{\circ}37^{\prime}$W) \\
Number of inputs & 2560 \\
Frequency range & 400 -- 800 \,MHz \\
Frequency resolution & 0.39\,MHz \\
Wavelength range & 75 -- 37 \,cm \\
Redshift range & z = 2.5 -- 0.8 \\
Epoch & 11 -- 8\,Gyr \\
E-W FOV & 2.5$^{\circ}$ -- 1.3$^{\circ}$ \\
N-S FOV & $\sim$90$^{\circ}$ about zenith \\
Angular resolution & 0.52$^{\circ}$ -- 0.26$^{\circ}$ \\ 
Spatial resolution & 10 -- 50\,Mpc\\\hline
\end{tabular}
\caption{The salient features of the CHIME instrument.}
\label{tab:salient}
\end{table}

As described in a companion paper in these proceedings\cite{bandura:2014}, we have built and are commissioning the CHIME Pathfinder consisting of two adjacent cylindrical dishes, each 20\,m wide by 37\,m long.  The Pathfinder serves as a test-bed for developing instrumentation and as a platform for testing calibration and analysis techniques. It will have sufficient sensitivity to measure BAO at low redshift. 

Due to the bright foreground signals present in the radio sky, accurate calibration techniques are critical if 21\,cm intensity mapping is to succeed.  In these proceedings we describe the calibration requirements for CHIME and our plans to meet them. We illustrate these plans with early Pathfinder commissioning data and with end-to-end simulation studies.

\section{Calibration Overview}
CHIME is a spatial interferometer: the data we record are correlations between signals from the feeds in the instrument. In this section, we present an overview of the quantities we must calibrate for CHIME, as well as the requirements on the precision of the calibration.

\subsection{Calibration Quantities}
The main task of CHIME calibration is to convert measured voltages from the correlator into sky maps with true sky values. We can write the frequency ($\nu$) dependent response of feed $i$ to an incident electric field $\varepsilon_a(\mathbf{\hat{n}},\nu)$ as\cite{shaw:2014b}
\begin{equation}
F_i(\phi,\nu) = \int d^{2}\mathrm{n}~g^{a}_{i}(\nu) A^a_i(\mathbf{\hat{n}}; \phi,\nu) \varepsilon_a(\mathbf{\hat{n}},\nu) e^{2\pi i \mathbf{\hat{n}} \cdot {\bf u}_i(\phi,\nu)},
\end{equation}
where $g_{i}^{a}(\nu)$ is the frequency-dependent complex receiver gain and $A^a_i(\mathbf{\hat{n}};\phi,\nu)$ is the antenna pattern of feed $i$ and polarization state $a$, and we implicitly sum over the two polarization states. The antenna pattern includes telescope pointing; because CHIME is a transit interferometer this is which is just the Earth's rotation angle $\phi$ and can simply calculated from sidereal time, and I will drop the explicit dependence. The measured voltage is referred to as a visibility and is given by the time-averaged correlation between the response of two feeds $i$ and $j$ ($i=j$ is referred to as an auto-correlation, and $i \ne j$ is a cross-correlation) and includes instrument noise $n_{ij}$: 

\begin{equation}
	V^\mathrm{meas}_{ij} = \langle F_{i}F_{j}^{*} \rangle = \int B_{ij} \langle \varepsilon(\mathbf{\hat{\mathrm{n}}},\nu) \cdot \varepsilon(\mathbf{\hat{\mathrm{n}}^{\prime}},\nu)  \rangle d^{2}\hat{\mathrm{n}}  + n_{ij}
\label{eqn:measuredvis}
\end{equation}

\noindent This can be decomposed into maps of the Stokes parameters $T(\mathbf{\hat{n}})$, $Q(\mathbf{\hat{n}})$, $U(\mathbf{\hat{n}})$, $V(\mathbf{\hat{n}})$, as
\begin{eqnarray}
V^{\mathrm{meas}}_{ij} = & \int d^2\mathbf{\hat{n}} \left[ B^T_{ij}(\mathbf{\hat{n}},\nu) T(\mathbf{\hat{n}},\nu) + B^Q_{ij}(\mathbf{\hat{n}},\nu) Q(\mathbf{\hat{n}},\nu) + B^U_{ij}(\mathbf{\hat{n}},\nu) U(\mathbf{\hat{n}},\nu)  + B^V_{ij}(\mathbf{\hat{n}},\nu) V(\mathbf{\hat{n}},\nu) \right] \nonumber \\
 & + n^{T,Q,U,V}_{ij}
\end{eqnarray}
where the beam transfer functions, $B^{T,Q,U,V}_{ij}$ encode all of baseline, polarization, and antenna pattern information about the instrument. For more details on the formalism, see Shaw et al.~\cite{shaw:2014b}. The BAO signal is a fluctuation in the temperature map only, but we must calibrate the full polarization dependence in the beam transfer function to prevent polarized foreground signal from leaking into the BAO temperature signal of  interest to CHIME. Thus, to convert measured voltages into maps, we must:
\begin{itemize} 
\item Determine the antenna pattern of each feed and polarization, $A^a_i(\mathbf{\hat{n}}; \nu)$, at each frequency.
\item Characterize the complex gain of each receiver as a function of time and frequency, $g_i(t,\nu)$.  
\item Account for cross-talk in the instrument over a range of lag times; schematically $\tilde{F_i}(t) = \sum_j \alpha_{ij} F_j(t-\tau_{ij})$.
\item Characterize the instrument noise, $n_{ij}(t)$ in Kelvin.
\end{itemize}

\subsection{Calibration Requirements}
\label{sec:cal_reqs}

CHIME must produce sensitive, foreground-cleaned maps of the sky on co-moving scales between the instrument resolution and a few times the BAO scale of 148\,Mpc.  To measure BAO from the data power spectrum, we require that the systematic errors from propagating foreground filtering, calibration, and other effects not dominate the statistical error. This requirement drives the CHIME calibration accuracy requirements, but is made particularly difficult to achieve by the presence of extremely bright astrophysical foreground signals, notably synchrotron emission from our own and external galaxies. As measured at 408\,MHz by Haslam et al.\cite{Haslam:1982}, foreground brightness can be as high as $\sim$700\,K near the Galactic center with typical mean signals of 10 -- 20\,K at high galactic latitudes.  Synchrotron emission typically falls with increasing frequency as $\nu^{-2.7}$, yielding high-latitude brightnesses of $\sim$2 -- 3\,K at 800\,MHz.  In contrast, the expected brightness fluctuations in the 21\,cm emission from BAO are $\sim$0.1\,mK rms, thus the foreground signals must be filtered by one part in $10^5$ to measure the 21\,cm fluctuations accurately. Foreground filtering is made tractable by the spectral smoothness of the synchrotron emission, which arises from cosmic-ray electrons spiraling in Galactic magnetic fields.  

Setting the calibration requirements for CHIME is a complex task that requires end-to-end simulations of the CHIME experiment, from simulating visibilities with realistic inputs to filtering and analyzing the data to probe the effects of various instrument properties on the fidelity of the simulated data streams.  We are undertaking a comprehensive simulation program to aid all aspects of the experiment, from hardware design choices to the setting of calibration requirements.   The requirements we summarize below highlight the most critical aspects of CHIME's calibration, informed largely by our simulation program.  Further details may be found in Shaw et al \cite{shaw:2014a,shaw:2014b}.

{\bf Beam response} - The requirements for the calibration precision of the frequency- and polarization-dependent antenna pattern are driven by two effects that introduce foreground spectral structure: (1) frequency dependence in the antenna beam converts angular structure in the foreground signal into spectral structure in any given visibility, and (2) polarized foreground signals undergo Faraday rotation in the Earth's ionosphere and the interstellar medium which introduces spectral structure in our visibilities. Since the magnitude of these effects depends on aspects of the foreground signal that are poorly known, we make pessimistic assumptions when setting these requirements.  We have an ongoing simulation program to study the effects in detail and we note a key result here.  In an end-to-end simulation where we perturb the full-width-half-maximum (FWHM) of each feed beam by a fixed amount that varies randomly from feed to feed, we find that each beam FWHM must be known to 0.1\% in order not to bias the derived power spectrum by more that its statistical uncertainty.  We continue to study the effects of other beam pattern errors in a similar manner.  We discuss our plans for meeting these requirements in \S \ref{sec:beams_main}. 

{\bf Complex gain} - Establishing relative complex gain (amplitude and phase) as a function of time is required to achieve relative brightness accuracy and to reliably combine multiple visibilities. We have propagated random errors in the gain through the analysis pipeline (including foreground filtering and power spectrum estimation), and conclude that random variations must be less than 1\% on 60\,s time scales.  This level limits bias in our power spectrum estimation to $\sim$10\% of the statistical uncertainty.  This applies independently to the real and imaginary parts of the gain, and is hence a de facto requirement on gain phase. We discuss our plans for meeting this requirement in \S \ref{sec:gain_main}. We will not include calibration to obtain the absolute sky brightness for CHIME because it is not critical to achieve our science goals. This is because the we will measure fluctuations in brightness generated by BAO, and so only differences in power between spatially separated regions of the sky are relevant for BAO power spectrum constraints. 

{\bf Cross-talk} - Coupling between channels in the detection path (cross-talk) can occur between cables, between feeds on the cylinder, and within the digitizer and correlator boards. Cross-talk from cables between the focal line and the correlator have values of better than $-$120\,dB (for the low-loss shielded coaxial cables on the focal line) and better than $-$70\,dB (for additional cables in the electronics building). Laboratory measurements of cross-talk within the digitizer boards produce estimates of $\sim-$50\,dB. We are working to understand and constrain the cross-talk between feeds on the cylinder using a portable vector network analyzer (VNA) to inject a signal into one feed and measure the response of neighboring feeds.

Cross-talk in our cross-correlation measurements adds both noise and signal with coefficients which are typically stable in time. The noise comes from cross-talk between the cross-correlation baselines and the auto-correlations, and appears in the cross-correlation time streams as a DC offset. Cross-talk between different cross-correlation products will leak signal from one baseline into the other. This will not generically be an additive constant offset, but will appear as signal at different phases, corresponding to non-main-beam pointing on the sky. These coefficients will have to be measured, then estimated and extracted from the data during analysis of the full data set. 

{\bf Instrument Passband} - The process of foreground filtering requires knowledge of our effective frequency bandpass to a part in $10^5$.  This requirement mixes knowledge of our gain and of our antenna response as a function of frequency.  It will be met by demanding that the brightest foreground regions in our maps simultaneously have a smooth frequency spectrum to a part in $10^5$. This must include the two previously mentioned effects (mode mixing and Faraday rotation) which introduce foreground spectral structure that must be accounted for in the filtering process. The degree to which beam effects complicate this approach is being actively studied at this time, so it is difficult to specify a bandpass requirement independent of a beam knowledge requirement.

We will describe the ongoing efforts to map the CHIME beams and understand the complex gain of CHIME throughout the rest of these proceedings. In \S~\ref{sec:beams_main} we discuss our current understanding of the CHIME beams through transits of bright sources, and discuss our plan for precisely mapping the polarized beams to the required accuracy through a technique we call pulsar holography. In \S~\ref{sec:gain_main} we discuss gain measurement and stabilization through a broadband calibration injection system and by exploiting the redundancy in the CHIME instrument for relative gain calibration.

\section{CHIME Beams}
\label{sec:beams_main}
The beam from an individual feed element on the CHIME cylinder will extend from the northern horizon to the southern horizon, with an east-west width determined by the cylinder diameter (1.1$^{\circ}$ -- 2.2$^{\circ}$ from 800 -- 400\,MHz). We expect the polarization response to vary through the beam shape. Because a single recorded voltage contains the integrated signal from the large CHIME beam, it is particularly challenging to precisely calibrate the CHIME polarized beam shape. We can use transits of bright sources for basic characterization, and are developing a method of pulsar holography to address these challenges and achieve the required 0.1\% precision.

\begin{figure}
   \begin{center}
   \begin{tabular}{c}
   \includegraphics[width=0.7\textwidth]{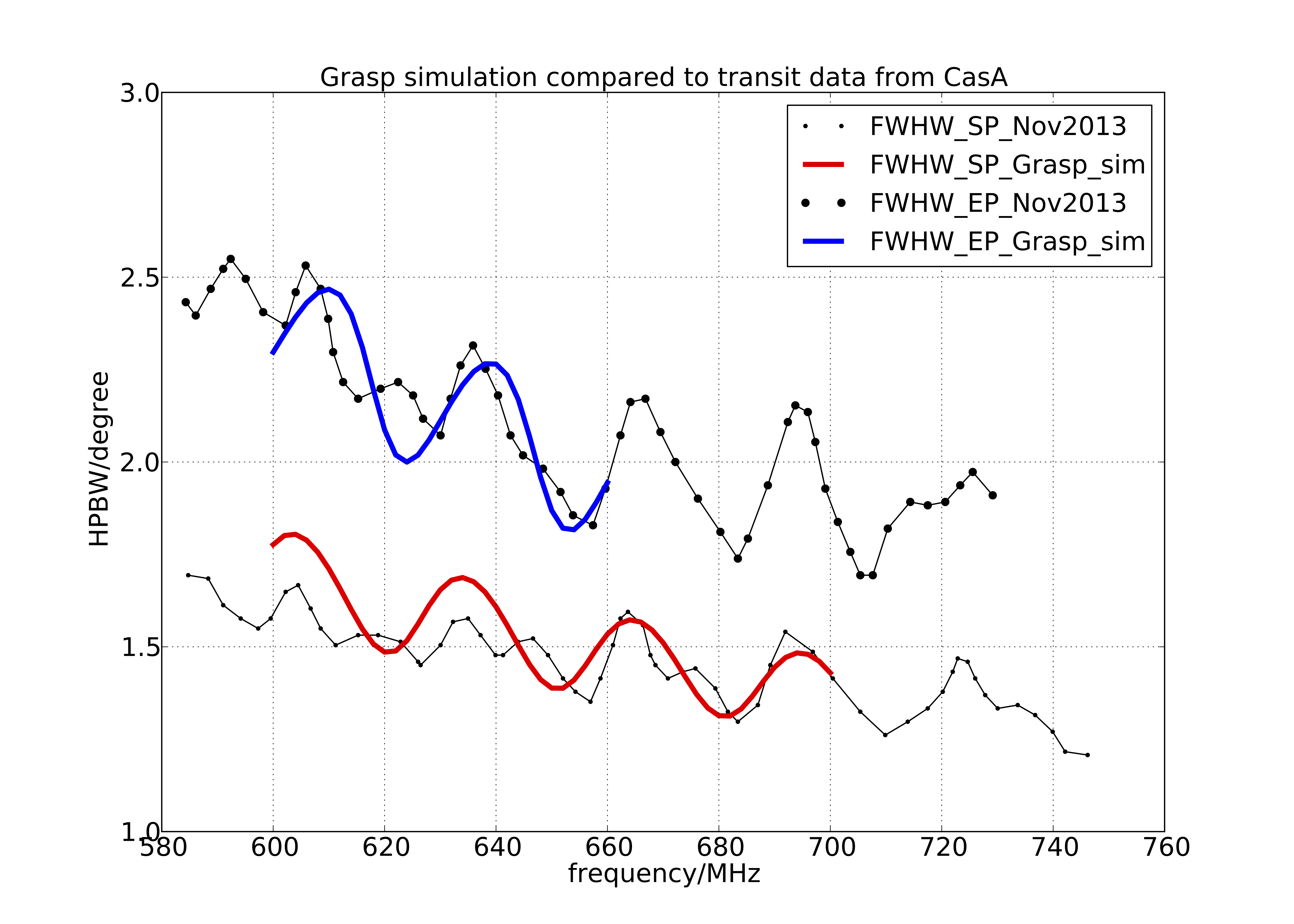}
     \end{tabular}
   \end{center}
   \caption{Shown is the fitted FWHM of the CHIME beam from a Cas A transit as a function of frequency in part of the CHIME instrument bandwidth. Data for the E-W (`EP') oriented beam and N-S (`SP') is in black. A 30\,MHz ripple is seen in the data at a frequency consistent with a standing wave between the cylinder and the feed ground plane. Simulations of the CHIME optics with the ground planes were performed, and the resulting FWHM behavior matches the shape of the modulation (red and blue curves).}
   \label{fig:transit_FWHM} 
 \end{figure} 
 
\subsection{Preliminary Beam Assessment with Source Transit Data}
We use daily transits of bright sources (CasA, CygA, TauA) through the CHIME beam to assess main-beam size and pointing as a preliminary step towards fully characterizing the instrument beam. We fit a Gaussian profile to the transit data to compute its FWHM and the peak response location in right ascension at the declination of the source. The fitted FWHM values from a single auto-correlation channel for all frequencies from a CasA transit are shown in Figure~\ref{fig:transit_FWHM}. There is a ripple at a frequency of $\sim$30\,MHz, corresponding to the characteristic frequency of a standing wave with wavelength 10\,m, approximately twice the distance between the focal line and the cylinder. These features are repeatable, suggesting that this is due to a true standing wave: some of the signal is reflecting off the ground planes of the feeds, bouncing off of the cylinder and back up to the feeds, and being detected again as signal. We modeled this system with the reflector modeling software GRASP\footnote{http:\/\/www.ticra.com\/products\/software\/grasp} and recovered a spectrum with the same peak locations seen in the data. These fits are included with the FWHM data in Figure~\ref{fig:transit_FWHM}. The declination-dependent location of the maximal response for different sources yields the orientation of each cylinder from the source transit data. Using source transits at different declinations, the reconstructed pointing is consistent with the 2$^{\circ}$ offset between grid north and celestial north at the DRAO. This 2$^{\circ}$ offset is a relic of the surveyor's usage of grid north when sighting the cylinder orientations and will not impact the pathfinder science result. 

\subsection{Beam Measurements from Pulsar Holography}
\label{sec:pulsarholo}
Transit data from bright, constant sources will not provide polarized beam response measurements at the required precision because it is difficult to disentangle the signal from other astrophysical sources, particularly those located in the far sidelobes. Pulsars provide a sky subtracted signal upon differencing `pulsar on' and `pulsar off' times. We will simultaneously measure pulsars with the stationary CHIME instrument and the tracking 26\,m telescope at the DRAO facility  to precisely map the polarized response of the CHIME beam. We attach a CHIME feed and amplifiers near the focus of the equatorially mounted 26\,m diameter dish. It has a focal ratio of 0.3, and we expect noise figures around 100\,K. Using a CHIME feed on a telescope with slightly larger focal ratio may lead to greater spillover and an increased thermal noise floor, which may be mitigated with a small cone around the feed. 100\,K noise temperatures are assumed for all calculations in this section. 

The signal from the 26\,m telescope will be fed directly to the CHIME correlator to be correlated with all of the CHIME channels. We refer to this measurement as `pulsar holography'. The technique of radio holography is a well-established method of mapping out a telescope's far-field antenna pattern. In this simplest case, radio holography requires two telescopes: a reference antenna that tracks a calibrator source (in this case the 26\,m telescope) and another whose beam is to be calibrated (in this case CHIME). Pulsar holography facilitates accurate beam mapping because we only measure correlated signals between the two telescopes: the pulsar and any background sky common to CHIME and the 26\,m telescope's beam. We difference `pulsar on' times and `pulsar off' times to remove the non-pulsed sky signal, leaving a clean measurement of the pulsar alone on fast time scales commensurate with the pulse period (of order 100\,ms) because neither the sky nor RFI varies on such short timescales. 

Since CHIME is a transit interferometer that sees each visible source once per day, in practice we will track pulsars in the range of $\pm15^\circ$ of the meridian with the 26\,m as they drift through CHIME's beams. This will generate a cross-correlation time-stream between the 26m and each of the CHIME feeds, as well as a 26\,m auto-correlation time stream. Together, they allow us to solve for the CHIME beams using the formalism we present below. 

Mapping the CHIME beams is done by solving for the Jones matrix for each dual-polarization CHIME feed, at every frequency, for as many points on the sky as possible given the spatial distribution of pulsars and the signal-to-noise of the instrument. A single transit of the brightest pulsar should allow us to learn about our primary beam to better than 1\% at that declination. In order to get a global beam fit we will need to find an optimal observing strategy in which we observe each pulsar for multiple transits. In addition, the transit of a single pulsar yields only one slice of the CHIME beam at the declination of the pulsar. To fill out the rest of the north-south beam, we must repeat this measurement for pulsars at a range of declinations. Pulsars are highly linearly polarized ($\sim15-90\%$) and they can be found at a wide range of declinations. 

Calibrating the CHIME beam is equivalent to determining the values of its Jones matrix. The instrumental Jones matrix relates the true electric field from the sky ($\mathbf{e}^{\mathrm{s}}$) to the measured electric field ($\mathbf{e}^{\mathrm{m}}$):

\begin{equation}
\mathbf{e}^{\mathrm{m}} = \mathbf{Je}^{\mathrm{s}}
\label{eqn:Jonesdef}
\end{equation}

The components of the electric field vectors are described by the two polarization states, in our case linear polarization ($\mathrm{e}_{X}$ and $\mathrm{e}_{Y}$), and $\mathbf{J}$ is a 2$\times$2 matrix which describes the instrument's response to the electric field: the diagonal components are transmission coefficients and the off-diagonals give the leakage between polarization states. The measured interferometric visibility is the time-averaged correlation between the electric field at two antennas, and can be expressed including the Jones matrices as a simplified version of Equation~\ref{eqn:measuredvis}:

\begin{equation}
V_{ij} =  \langle \mathbf{e}^{\mathrm{m}}_{i} \mathbf{e}^{\mathrm{m} \dagger}_{j} \rangle = \mathbf{J}_i \langle \mathbf{e}^{\mathrm{s}} \mathbf{e}^{\mathrm{s} \dagger} \rangle \mathbf{J}_j^\dagger
\end{equation}

\noindent Noting that  $\langle \mathbf{e}^{\mathrm{s}} \mathbf{e}^{\mathrm{s} \dagger} \rangle$ is just the true sky visibility (`coherency matrix': $\mathbf{P}(t)$), we can write the holographic response of the CHIME$\times$26m cross-correlation as

\begin{equation}
\mathbf{V}_{\mathrm{hol}}(t) =  \mathbf{J}_i (\hat{\mathbf{n}}(t)) \, \mathbf{P}(t)\, \mathbf{J}_{26}^\dagger(\hat{\mathbf{n}}(t))  + \mathbf{N_{\mathrm{hol}}}(t)
\label{eqn:Hol1}
\end{equation}

\noindent where $\mathbf{J}_{26}$ is the known, stable Jones matrix of the 26\,m telescope, $\mathbf{N_{\mathrm{hol}}}(t)$ is the measurement noise covariance matrix, and $\mathbf{J}_i(\hat{\mathbf{n}})$ is the  Jones matrix of a CHIME antenna $i$ as a function of position on the sky $\hat{\mathbf{n}}$. The latter is the quantity we want to solve for using the cross-correlation measurement between the 26\,m and CHIME. Assuming a known $\mathbf{J}_{26}$, we can use the measured autocorrelation from the 26\,m:

\begin{equation}
\mathbf{V}_{26}(t) =  \mathbf{J}_{26} \mathbf{P}(t) \mathbf{J}_{26}^\dagger  + \mathbf{N}_{26}(t)
\end{equation}

\noindent to rearrange and find an estimate for the pulsar's polarization matrix $\hat{\mathbf{P}}(t) = \mathbf{J}_{26}^{-1} \mathbf{V}_{26}(t) \mathbf{J}_{26}^{\dagger -1}$ which we can insert into Equation \ref{eqn:Hol1} to obtain,

\begin{eqnarray}
\hat{\mathbf{J}}_i (\hat{\mathbf{n}}(t)) & = \mathbf{V}_{\mathrm{hol}}(t) \mathbf{J}_{26}^{\dagger -1} \langle ( \mathbf{J}_{26}^{-1} \mathbf{V}_{26}(t) \mathbf{J}_{26}^{\dagger -1} \rangle )^{-1} 
\\
&= \mathbf{V}_{\mathrm{hol}}(t) \mathbf{V}_{26}(t) ^{-1}\mathbf{J}_{26}
\label{eqn:Hol2}
\end{eqnarray}

\noindent Because the noise covariance term has been neglected, this estimator is suboptimal. Although it is beyond the scope of this paper, one can show the bias introduced by the non-zero mean of $\mathbf{N}_{26}(t)$ can be estimated.

\begin{figure}
   \begin{center}
   \begin{tabular}{c}
   \includegraphics[width=0.7\textwidth]{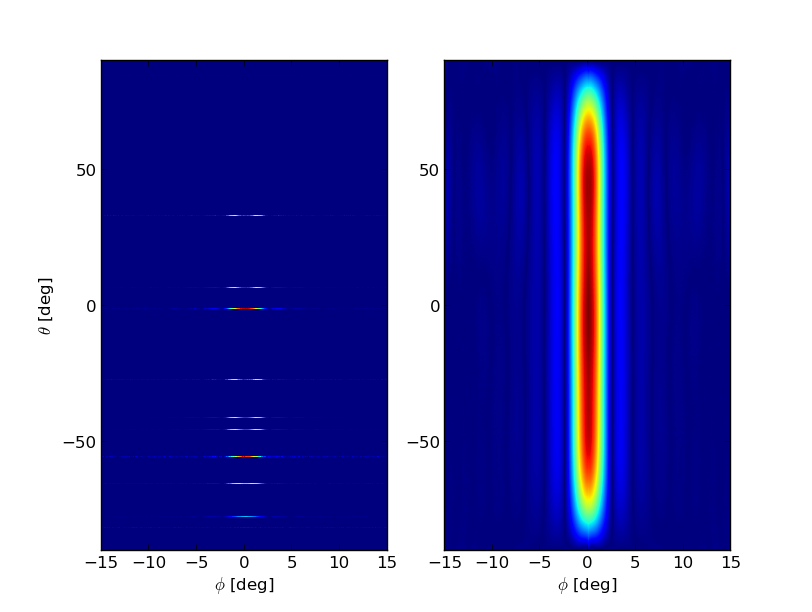}
     \end{tabular}
   \end{center}
   \caption{\textit{Left:}  Pulsar tracks through a simulated CHIME beam for 10 of the brightest pulsars from the ATNF pulsar catalog. \textit{Right:} The extrapolated CHIME beam from pulsar measurements, assuming a smooth and symmetric profile. A prior for the true profile will be informed from optics simulations. An example of an interference pattern for a source (CasA) between the 26\,m and CHIME is shown in Figure\ref{fig:casa_holo}}.
   \label{fig:pulsar_sims} 
 \end{figure}

Using Equation \ref{eqn:Hol1} we simulate a holographic visibility time stream and reconstruct a CHIME beam given a set of pulsar observations. This simulation was performed assuming a beam taken from Shaw et al.~\cite{shaw:2014b} and modeling the pulse-to-pulse flux variation as a log-normal distribution to generate the 26\,m auto-correlation timestream and the cross-correlation. We use these simulated time streams to find the Jones matrix via Equation \ref{eqn:Hol2}. We used the brightest sources from the ATNF pulsar catalogue ~\cite{ATNF:2005} which gave us a set of tracks through our beam at different declinations. The tracks and resulting beam (assuming it is both smooth and symmetric) is shown in Figure~\ref{fig:pulsar_sims}. Work is currently underway to fit these tracks with a set of basis functions to parametrize the full beam shape. The choice of basis functions and the fit will require both pulsar measurements and some guidance from optical modeling.

The above calculations assume a known Jones matrix for the 26\,m telescope. The 26\,m telescope is equatorially mounted such that the telescope rotates with parallactic angle as it tracks the source in the sky. Equatorial tracking makes calibrating the polarized response of the 26\,m difficult. This is because the least error-prone polarization calibration is derived from tracking polarized sources as they rotate with parallactic angle, and fitting for instrumental polarization as the term that is constant with parallactic angle rotation. To properly calibrate the 26\,m, we will require a large telescope which is alt-azimuth mounted. We plan to use the 46\,m dish at the Algonquin Radio Observatory (ARO) to perform a very long-baseline interferometric (VLBI) measurement. We will first calibrate the alt-az 46\,m telescope with a stable, bright pulsar. We then perform the VLBI between the 26\,m DRAO telescope and the 46\,m ARO telescope. The solution for obtaining the 26\,m calibration from the ARO telescope is identical to the expressions above for the 26\,m$\times$CHIME beam calibration. There are a few additional complications since we cannot directly cross-correlate the two VLBI telescopes, and we must include the effect of the atmosphere, particularly Faraday rotation as the source photons propagate through the Earth's ionosphere. 

\begin{figure}
   \begin{center}
   \begin{tabular}{c}
   \includegraphics[width=0.9\textwidth]{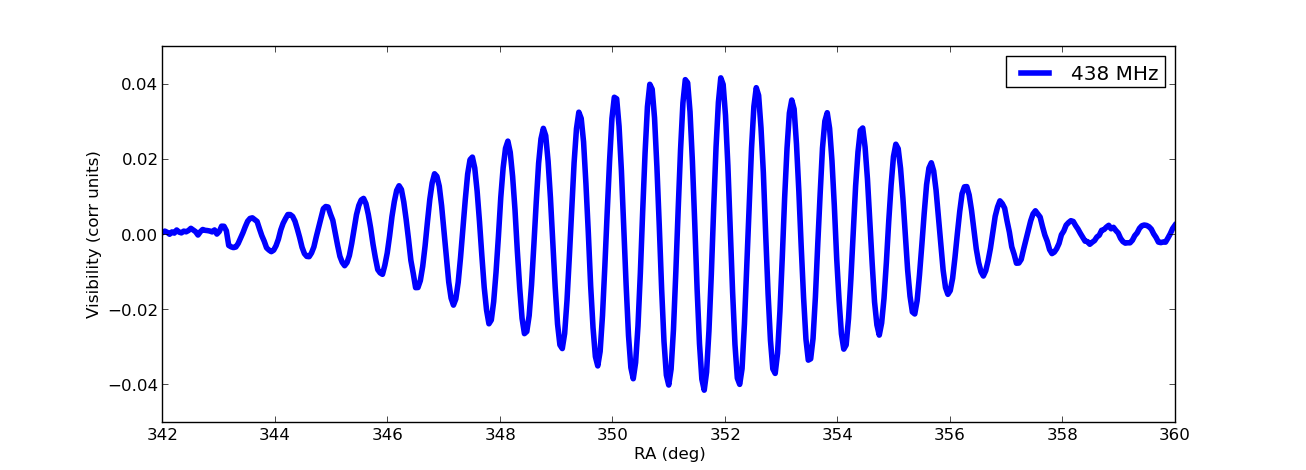}
     \end{tabular}
   \end{center}
   \caption{Interference fringes between the tracking 26\,m telescope at DRAO and CHIME as CasA transits through the CHIME beam, at $\nu$=438\,MHz. }
   \label{fig:casa_holo} 
 \end{figure}

We began pulsar calibration measurements between the 26\,m and CHIME in May 2014. Analysis of the signal-to-noise from the measurements and initial beam measurements from pulsars is underway, but we began by tracking CasA with the 26\,m as it transited through the CHIME beam. The first fringes from this transit are shown in Figure~\ref{fig:casa_holo} for frequency $\nu=438$\,MHz.

\section{Gain Determination}
\label{sec:gain_main}
The complex gain of each receiver depends on frequency and changes with time.  Gain variations are typically attributed to $1/f$ noise in the amplifiers and to thermal changes in the environment.  To mitigate systematic errors, we must determine and apply gain corrections on time scales faster than the rate at which the sky signal changes. In this section we describe various techniques we plan to use for gain calibration on CHIME, including laboratory characterization, broadband noise injection, switched sky sources (pulsars), and redundant baseline modelling.

\subsection{Thermal characterization}
\label{sec:gain_char}
We have measured LNA gain of several CHIME amplifiers in a temperature-controlled chamber to find their thermal susceptibility over the range $-20$\,C to $+50$\,C.  The results for several of these amplifiers indicate that the typical amplifier gain varies by about $-$0.05\% per Kelvin.  The CHIME amplifiers are equipped with thermal sensors that allow us to monitor temperatures on time scales faster than they will typically vary.  In addition, we will occasionally terminate selected receivers with temperature-stabilized loads to (re)characterize the thermal susceptibility in the field.   We can use these data to correct for temperature dependent gain drifts in post-processing.

\subsection{Broadband Calibration Injection Signal}
\label{sec:bis}
Using an injected source to measure relative receiver gains through time is a well-established practice in radio astronomy. We have implemented a Broadband Injection Signal (BIS) for gain calibration on the pathfinder across the full CHIME bandwidth. A description of the BIS apparatus, the analysis technique, and preliminary results from tests on the pathfinder are presented in this section. 

\subsubsection{The BIS Apparatus}
\begin{figure}
   \begin{center}
   \begin{tabular}{c}
   \includegraphics[width=0.7\textwidth]{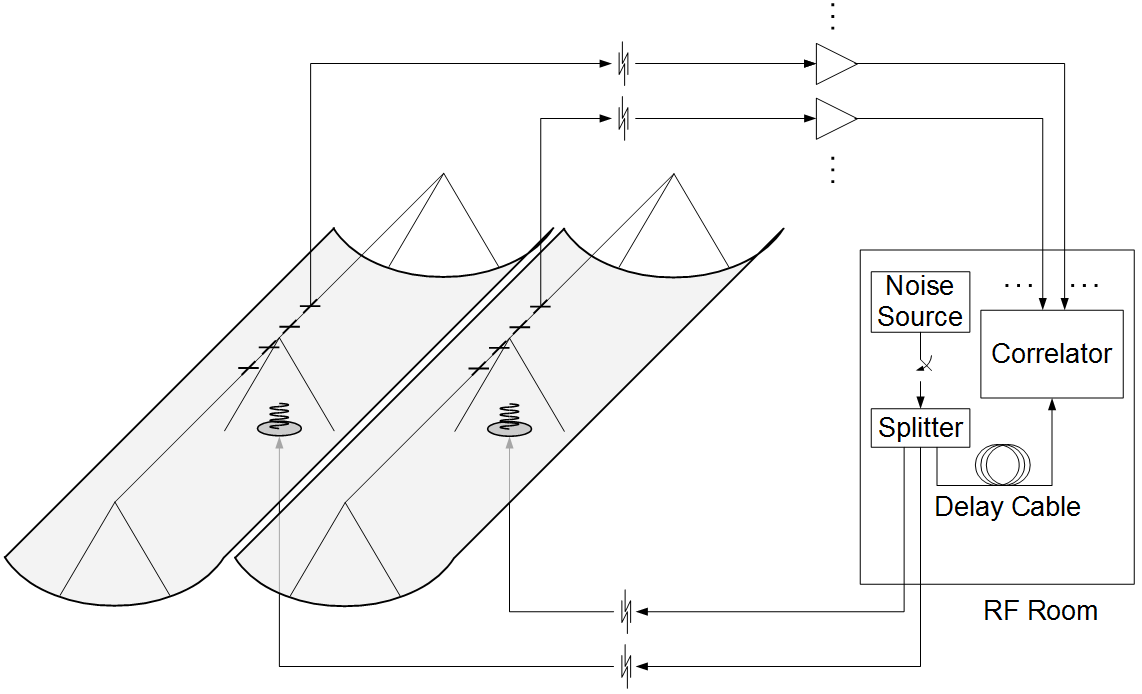}
   \end{tabular}
   \end{center}
   \caption[noise_injection_setup] 
   { \label{fig:noise_injection_setup} 
	The setup for the broadband spectrum signal injection on the CHIME pathfinder. Inside the electronics building, a switched
	 noise source generates the reference signal. A copy of this signal is sent to
	the correlator through a coaxial delay cable.  Two additional coaxial cables bring two copies of the reference 
	signal to helix antennas located at the base of each cylinder. }
   \end{figure}

A diagram of the setup for injecting the broadband calibration signal is shown in Fig.~\ref{fig:noise_injection_setup}. Inside a screened room a switched noise source is installed. The output of this source is filtered to impose the CHIME passband and then passed through a three-way splitter. Two of the three splitter outputs are connected to coaxial cables which bring the calibration signals to the base of the two cylinders. They each terminate in a helix antenna located at the bottom of the CHIME cylinders, which injects a circularly polarized signal into the CHIME feeds. The third port of the splitter is connected directly to the CHIME correlator through a delay cable. The delay cable is required to ensure that the injected signals arrive at the channelizer well within the FFT ($\sim$2\,$\mu$s)length for channelizing into 400 -- 800\,MHz. 

The attenuation of the reference channel is adjusted so that its level at the input of the Analog to Digital 
Converter (ADC) minimizes the noise penalty due to quantization and ADC noise. The attenuation of 
the BIS signals is adjusted so the additional noise contribution is $\sim$50\,K, roughly doubling the CHIME system noise temperature.  This ensures that the combined signal level from the sky and BIS remains higher than our quantization noise but low enough that the amplifiers and ADCs remain in their linear signal regime \cite{Mena13,Davis13}. We are currently working to implement a gating scheme in the CHIME correlator which will allow us to reduce the time we have doubled our noise with BIS to short, occasional bursts.

\subsubsection{BIS results}
We demonstrate this strategy with a small data set (18\,minutes of data with 4 channels from each cylinder). We are continuing to improve the system and in particular working to choose the period and duration of the injected noise, but the preliminary results are encouraging: we can already calibrate to $\sim$1\% with the current injection setup. 

The BIS signal has a Gaussian distribution with zero mean and is broadband, allowing us to calibrate the entire CHIME bandpass simultaneously. The resulting measured visibility is a version of Equation~\ref{eqn:measuredvis} with gain terms only. The measured visibilities $\mathbf{V}^{\mathrm{meas}}$ for BIS on and BIS off are:

\begin{eqnarray}
\mathbf{V}^{\mathrm{meas}}_{\mathrm{on}} & = & \mathbf{g}\mathbf{g}^{\mathrm{H}}\sigma_s^2 + \mathbf{G}\mathbf{V}^{\mathrm{sky}}\mathbf{G}^{\mathrm{H}}+\mathbf{N}_{\mathrm{on}}
\\
\mathbf{V}^{\mathrm{meas}}_{\mathrm{off}} & = &  \mathbf{G}\mathbf{V}^{\mathrm{sky}}\mathbf{G}^{\mathrm{H}} +\mathbf{N}_{\mathrm{off}} 
\end{eqnarray}

\noindent where $\mathbf{V}^{\mathrm{sky}}$ is the $N \times N$ matrix containing the true sky complex visibility, $\mathbf{G}$ is the $N \times N$ matrix with complex receiver gains ($\mathbf{g}$) on the diagonal, and $\sigma_{s}$ is the BIS signal level
(the direction dependent array response for the BIS is assumed to be known and is not shown explicitly).
The two receiver noise terms $\mathbf{N}_{\mathrm{on}}$ and $\mathbf{N}_{\mathrm{off}}$ will not in principle be exactly identical but their difference is a white noise term and so decreases as $1/ \sqrt{\tau \Delta\nu}$ as long as both the sky and receiver response remains unchanged during observations. With an accurate measurement of $\sigma_{s}$ from the auto-correlation of the noise source channel, the difference between the measurement with BIS on and BIS off, divided by $\sigma_{s}^{2}$ is:
\begin{equation}
\frac{\Delta \mathbf{V}^{\mathrm{meas}}}{\sigma_{s}^{2}} = \mathbf{g}\mathbf{g}^{\mathrm{H}} + \frac{\Delta \mathbf{N}}{\sigma_{s}^{2}}
\end{equation}

\noindent yielding a quantity proportional to the receiver gain with a noise bias. In practice, the data for each pulse (BIS on, BIS off) is separately averaged and the mean values are differenced. This produces a data set of differenced values at half the cadence of the switching, in this case 10\,s (5\,s on, 5\,s off, with a data sample time of 84\,ms). The differenced time stream is a noisy estimator for the gain solution, which can be obtained by minimizing the residual noise in the differenced data set. This can be efficiently solved by formulating this as a rank-1 approximation problem (singular value decomposition, SVD), and for each time bin we solve the following to find an estimate of the length-$N$ gain vector $\mathbf{\hat{g}}$:

	\begin{equation}
 \mathbf{\hat{g}} =\underset{\mathbf{g}}{\hbox{argmin}} \left\|  \frac{\Delta \mathbf{V}^{meas}}{\sigma_{s}^{2}} -\mathbf{g}\mathbf{g}^{\mathrm{H}} \right\|_F^2=  \sqrt{\lambda_{\mathrm{max}}}\mathbf{u}_{\mathrm{max}}
	\label{eqn:rank_one}
	\end{equation}

\noindent where $\lambda_{\mathrm{max}}$ is the largest eigenvalue for each channel and $ \mathbf{u}_{\mathrm{max}}$ is its corresponding eigenvector in the eigenvalue
decomposition of $\Delta \mathbf{V}^{\mathrm{meas}}$. 

\begin{figure}[h]
   \begin{center}
   \begin{tabular}{c}
   \includegraphics[width=0.9\textwidth]{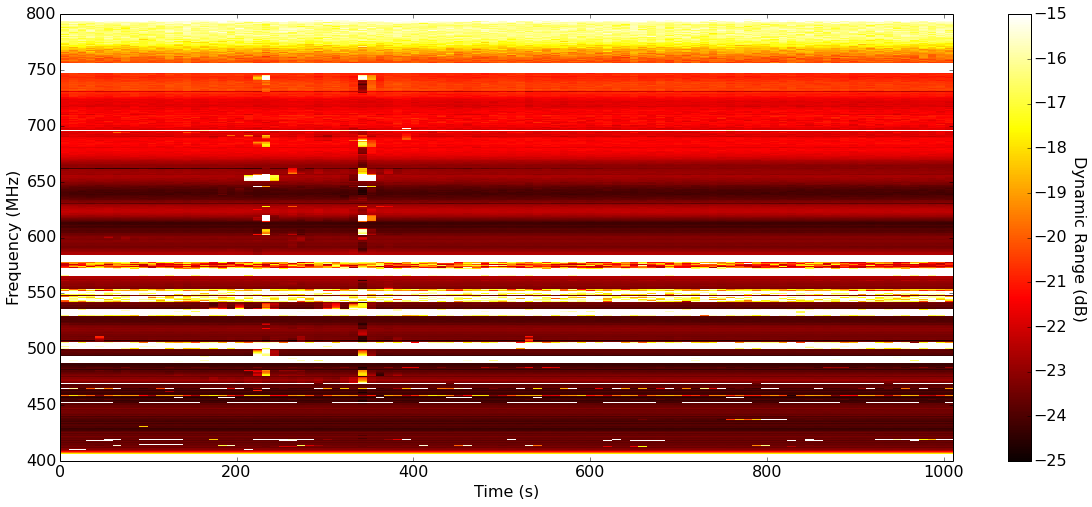}
   \end{tabular}
   \end{center}
   \caption[dynamic_range] 
   { \label{fig:dynamic_range} 
      The ratio of the first and second eigenvalues in the eigenvalue decomposition of $\Delta \mathbf{V}$ (`dynamic range') obtained during the first tests of the white spectrum signal injection. The dynamic range determines the accuracy with which we can 
recover the receiver gains and is $-$20 -- 25\,dB
in most of the CHIME band, corresponding to relative gain errors below 1\%. Light $\sim$ 6\,MHz wide horizontal stripes,
where the performance is poor, correspond to TV channels in the CHIME band.}
   \end{figure}

It can be shown that the error on the estimated gain is determined by the ratio between the first and second largest eigenvalues in Equation~\ref{eqn:rank_one}. We call this the dynamic range, and it should be $<-20$\,dB to achieve 1\% relative gain calibration precision. The dynamic range for the BIS measurement is shown in Figure~\ref{fig:dynamic_range}, and in these preliminary measurements is $\sim-$20\,dB for most of the CHIME band. 

\begin{figure}
   \begin{center}
   \begin{tabular}{c}
   \includegraphics[width=0.9\textwidth]{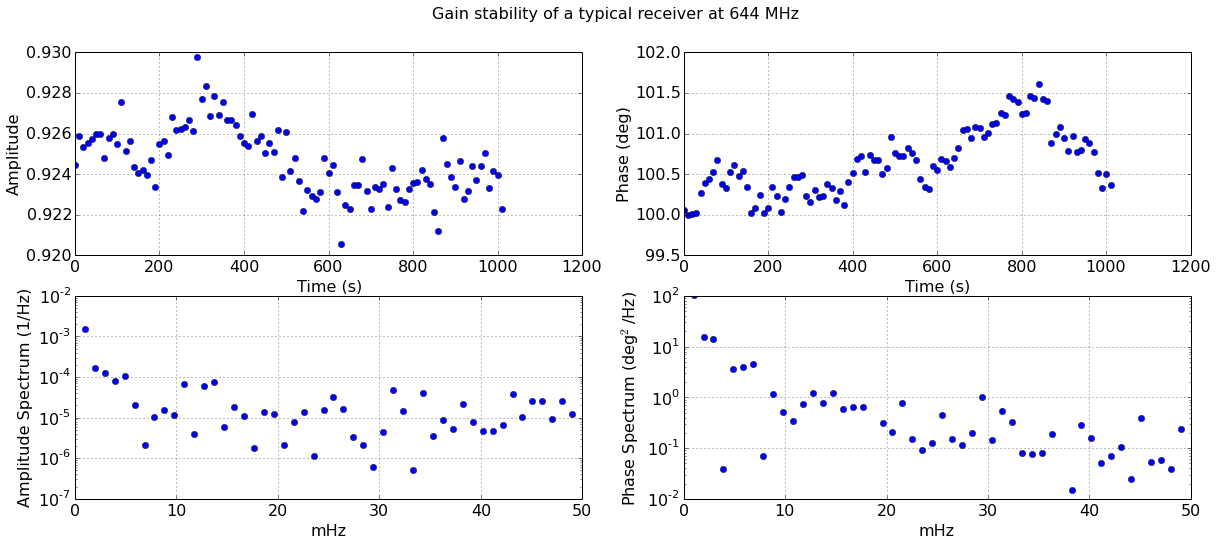}
     \end{tabular}
   \end{center}
   \caption{Gain stability (both amplitude and phase) performance of a typical receiver ($\nu =644$ MHz). \textit{Top:} Relative gain amplitude and phase during the 18\,minute noise injection data file. \textit{Bottom}: Power spectrum of the relative gain amplitude and phase from the time stream. To fit the power spectrum and derive stability, we will take additional data with with longer files and faster switching. }
   \label{fig:ni_gain} 
 \end{figure}

These measurements can also be used to assess the gain stability (both amplitude and phase) over time. Fig.~\ref{fig:ni_gain} shows the gain stability performance of a typical receiver for the frequency centered at 644\,MHz from preliminary data. This is a work in progress and so far we do not have enough data to accurately assess the time scales on which we must calibrate the gain.

We will apply the gains obtained from BIS before further analysis, such as co-adding data or averaging data from other visibilities at the same baseline distance. 

\subsection{Gain Calibration with Pulsars}
\begin{figure}
   \begin{center}
   \begin{tabular}{c}
   \includegraphics[width=0.9\textwidth]{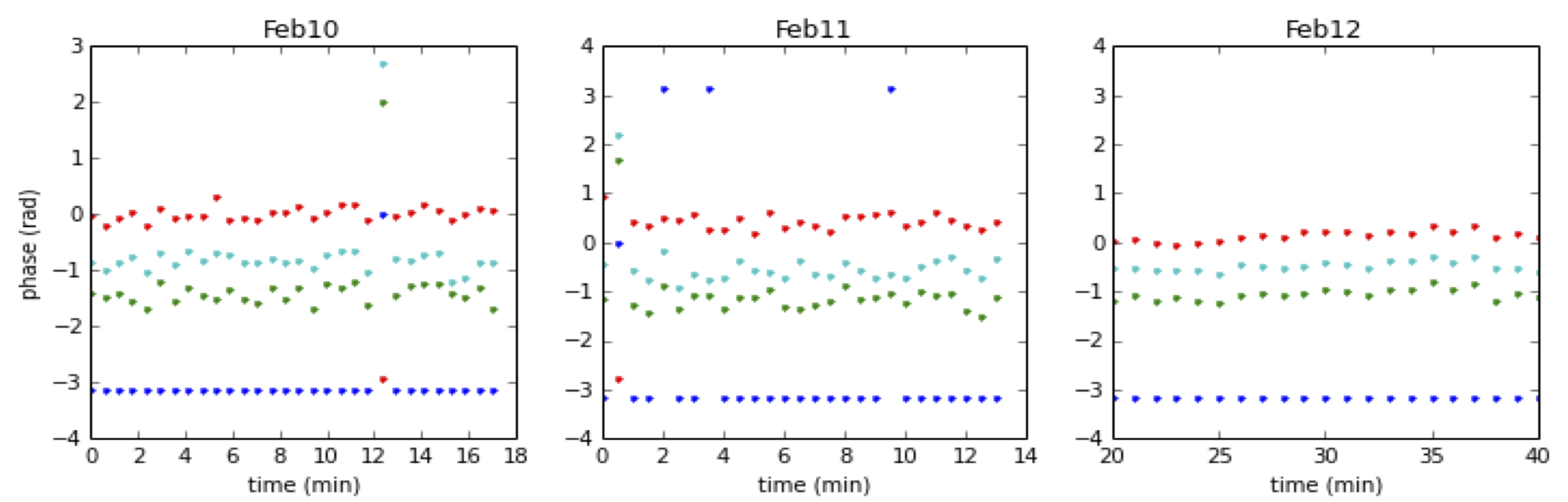}
   \end{tabular}
   \end{center}
   \caption[phase_stability] 
   { \label{fig:phase_stability} 
      Relative phase stability averaged across frequencies during a pulsar transit over three days for three feeds on a single cylinder referenced to the phase of a fourth. This fiducial channel (blue) has been set to $-\pi$. The data is preliminary and include variation between pulses which are intrinsic to the pulsar. The purpose of the data was primarily to understand timing, cable delays, and other instrumental properties but is still an initial step towards understanding the gain stability using pulsars.}
   \end{figure}

We have observed pulsar transits with CHIME alone. These function as a switching noise source much like BIS. The formalism and gain estimation techniques from Section~\ref{sec:bis} apply to pulsars as well as the noise source, and can be used to extract relative gains between channels. Using a series of fast-cadence pulsar data from CHIME over the course of three days, the complex gain (amplitude and phase) stability were assessed. The resulting phase stability, relative to a single channel, is shown in Figure~\ref{fig:phase_stability}. The data is also preliminary and cannot yet assess stability from this measurement.

\subsection{Redundant Baselines}

The large number of redundant baselines within CHIME may be exploited for calibration purposes.  For an array with $N$ feeds there will be $N^2$ measured visibilities. If the primary beams of each feed are identical, the true-sky visibilities $V^{\mathrm{s}}_{ij}$ will be common to all measurements with the same baseline separation and can be written as a simplified version of Equation~\ref{eqn:measuredvis}: $V^{\mathrm{meas}}_{ij} = g_i g^*_j V^{\mathrm{sky}}_{ij}$, with the same baseline separation ${\bf u}_{ij}$. This creates an overdetermined system that can be solved simultaneously for the complex gain, $g_i$, and the true-sky visibilities independent of any ancillary data up to an overall calibration scale factor and sky tilt \cite{noordam1982, yang1988, wieringa1992}.

Several different algorithms have been developed to solve redundant baseline systems \cite{yang1988,wieringa1992,liu2010,wijnholds2012,marthi2014} and the method has been tested on existing arrays \cite{noordam1982, noorishad2012}.   We have developed an efficient, hybrid method for solving the system: to solve for the amplitude of the complex gain, we build on the work of Liu et al\cite{liu2010} and linearize the problem by taking the logarithm of the system.  To solve for the phase, we use an SVD technique applied to the difference,  $V^{\mathrm{meas}}_{ij} - g_i g^*_j V^{\mathrm{sky}}_{ij}$, to find the best rank-1 approximation to the gain matrix $g_ig^*_j$\cite{Sitwell2014}.  In simulations of a 12-feed instrument observing a realistic sky signal with forecast instrument noise and identical primary beams, we recover the input relative gain amplitude without statistical bias to a precision of better than 0.1\% for each 10\,s observation and 0.39\,MHz frequency band.  In the same simulation, we recover the gain phase with an rms uncertainty of 0.7$^{\circ}$ per observation and frequency band, where the requirement is $\sim$0.01\,rad (0.6$^\circ$) and is included in the simulations to derive a calibration requirement on complex gain.

In a real instrument, the primary beam of each feed will differ slightly from the average, which leads to non-redundant visibilities even if the baselines are redundant.  To see how important such effects are for our calibration determination, we produced a simulation in which we perturb the FWHM of each primary beam by 2\% (rms) and solve the redundant baseline system ignoring the beam variations.  In this case, our recovered gain amplitude has a systematic error of 0.2\%, which exceeds its 0.1\% statistical uncertainty.

We are investigating extensions of the above algorithm that can accommodate primary beam variations and have had some success with their early implementation.  Our approach is to expand the primary beam in a perturbative expansion and to simultaneously solve for the sky signal, the complex gain, and the beam expansion coefficients.  To date, we have treated the beam as a 2-term expansion in Hermite polynomials and have applied this technique to the perturbed beam simulation described above.  In that test, we were able to recover the input gain without statistical bias; the rms uncertainty was $\sim$20\% larger than the idealized redundant algorithm, but it was still less than 0.1\% per observation and frequency band.  We continue to develop this complementary approach to the calibration of CHIME.

\section{Conclusion}
We have developed requirements for CHIME calibration and formed a plan to calibrate the polarized beams and complex gains of the CHIME instrument. We have a simulation pipeline currently in place that has already provided valuable specifications for the beam and gain calibration requirements. We are working on using it to assess acceptable levels of cross-talk, aliasing, and instrumental passband, particularly where they may inform instrument design such as feed spacing. 

We will have made the first pulsar holography measurements to map the polarized beams in May 2014. We have implemented a basic noise injection system which has proven to calibrate complex gains to 1\%, our target for controlling gains on short time scales. We have also confirmed that our system noise is on-target. Further progress on improvements to the noise injection system, calibration of the 26\,m dish at the DRAO for pulsar holography, and testing redundant baseline algorithms on data will be done in the near future. 

\acknowledgments     

We are very grateful for the warm reception and skillful
help we have received from the staff of the Dominion Radio
Astrophysical Observatory, operated by the National Research Council Canada.

We acknowledge support from the Canada Foundation for Innovation, 
the Natural Sciences and Engineering
Research Council of Canada, the
B.C. Knowledge Development Fund,  le Cofinancement gouvernement du
Qu\'ebec-FCI, the Ontario Research Fund, the CIfAR Cosmology and
Gravity program, the Canada Research Chairs program, and the National
Research Council of Canada. 
M.~Deng acknowledges a MITACS
fellofowship.  
 We thank Xilinx  and the XUP for their generous donations.  

\bibliography{SPIE_calbib}   
\bibliographystyle{spiejour}   
\end{document}